\documentclass[12pt]{article}
\usepackage{enumerate}
\usepackage{amsfonts}
\usepackage{amsmath}
\usepackage{amssymb}
\usepackage{amsthm}
\usepackage{latexsym}
\usepackage{color}

\def\A{\mathcal{A}}
\def\B{\mathcal{B}}
\def\C{\mathcal{C}}
\def\H{\mathcal{H}}
\def\K{\mathcal{K}}

\def\S{\mathfrak{S}}
\def\C{\mathfrak{C}}
\def\T{\mathfrak{T}}
\def\B{\mathfrak{B}}

\newcommand{\rank}{\mathrm{rank}}
\newcommand{\id}{\mathrm{Id}}
\newcommand{\Tr}{\mathrm{Tr}}

\newcommand{\shs}{\hspace{1pt}}
\newcounter{defin}  \newcounter{lemma}  \newcounter{theorem}
\newcounter{property} \newcounter{corol}  \newcounter{remark} \newcounter{example}

\newenvironment{theorem}{\par\refstepcounter{theorem}     \textbf{Theorem \thetheorem.}\ }{\rm\par}
\newenvironment{property}{\par\refstepcounter{property}     \textbf{Proposition \theproperty.}\ }{\rm\par}
\newenvironment{corollary}{\par\refstepcounter{corol}     \textbf{Corollary \thecorol.} }{\rm\par}

\newenvironment{remark}{\par\refstepcounter{remark}     \textbf{Remark \theremark.}}{\rm\par}
\newenvironment{example}{\par\refstepcounter{example}     \textbf{Example \theexample.}}{\rm\par}

\begin{document}

\title{On quantum channels and operations preserving finiteness of the von Neumann entropy}
%\title{Universal tight continuity bounds for characteristics of energy-constrained quantum systems and channels}
\author{M.E.Shirokov\footnote{Steklov Mathematical Institute, email:msh@mi.ras.ru}, A.V.Bulinski\footnote{Moscow Institute of Physics and Technology, email:andrey.bulinski@yandex.ru}}
\date{}
\maketitle
%\vspace{50pt}
\begin{abstract}
We describe the class (semigroup) of quantum channels  mapping states with finite entropy into states with finite entropy.
We show, in particular, that this class is naturally decomposed into three convex subclasses, two of  them are closed under concatenations and tensor products. We obtain asymptotically tight universal continuity bounds for the output entropy of two types of quantum channels:
channels with finite output entropy and energy-constrained  channels preserving finiteness of the entropy.
\end{abstract}

\tableofcontents

\section{Introduction}

The output entropy $H_{\Phi}(\rho)\doteq H(\Phi(\rho))$  of a quantum channel $\Phi$ is an important characteristic of this channel essentially used in analysis of its information abilities \cite{H-SCI,Wilde}. For an arbitrary channel $\Phi$ between quantum systems $A$ and $B$ the output entropy $H_{\Phi}$ is a lower semicontinuous concave function on the set $\S(\H_A)$ of all states of the input system $A$ taking values in $[0,+\infty]$. If the function $H_{\Phi}$ is finite for all input states then it is bounded and continuous on the set $\S(\H_A)$ \cite[Theorem 1]{OE}.

In contrast to many other characteristics of quantum channels (such as the mutual and coherent informations, the constrained Holevo capacity, etc., see \cite{H-Sh-2}) the output entropy $H_{\Phi}(\rho)$ may take infinite values at states $\rho$ with finite entropy (the simplest example is the channel transforming all input states into a single state with infinite entropy). Nevertheless, there is a large class (semigroup) of quantum channels
mapping states with finite entropy into states with finite entropy, i.e. such channels $\Phi$ that
\begin{equation}\label{PFE}
H(\rho)<+\infty\quad \Rightarrow \quad H_{\Phi}(\rho)<+\infty.
\end{equation}
It is shown in \cite{PCE} that any channel $\Phi$ possessing property (\ref{PFE})
preserves local continuity of the entropy: if a sequence $\{\rho_n\}$ converges to a state $\rho_0$ then
\begin{equation}\label{PCE}
\lim_{n\to+\infty}H(\rho_n)=H(\rho_0)<+\infty\quad \Rightarrow \quad\lim_{n\to+\infty}H_{\Phi}(\rho_n)=H_{\Phi}(\rho_0)<+\infty.
\end{equation}

In this paper we describe the structure of the semigroup of all quantum channels possessing the
equivalent properties (\ref{PFE}) and (\ref{PCE}). We show, in particular, that this semigroup is naturally decomposed into three
classes $\A$, $\B$ and $\C$. Classes $\A$ and $\B$ are convex  and closed under concatenations and tensor products. The class $\C$ contains channels of the form $\lambda\Phi+(1-\lambda)\Psi$, $\lambda\in(0,1)$, where $\Phi$ and $\Psi$ are channels from the classes $\A$ and $\B$ correspondingly. The class $\C$ is convex but
not closed under concatenations and tensor products. Channels from this class are characterised by existence of input state at which
both the output entropy and the entropy exchange are infinite. \smallskip

It is well known that the von Neumann entropy is continuous on the set
$$
\mathfrak{C}_{H_A,E}=\left\{\rho\in\mathfrak{S}(\mathcal{H}_A)\,|\,\mathrm{Tr} H_A\rho\leq E\right\}
$$
of states with mean energy not exceeding $E$ if (and only if) the Hamiltonian  $H_A$ of system $A$ satisfies  the Gibbs condition
\begin{equation*}%\label{H-cond}
  \mathrm{Tr}\, e^{-\lambda H_{A}}<+\infty\quad\textrm{for all}\;\lambda>0.
\end{equation*}
Under this condition the set $\mathfrak{C}_{H_A,E}$ is compact \cite{H-c-w-c}. So, in this case the equivalence of (\ref{PFE}) and (\ref{PCE})
implies that the output entropy of any channel $\Phi$ satisfying (\ref{PFE}) is uniformly continuous on the set
$\mathfrak{C}_{H_A,E}$ for any $E$. In this paper we obtain uniform continuity bound for the function $\rho\mapsto H_{\Phi}(\rho)$ on the set
$\mathfrak{C}_{H_A,E}$ in explicit form. This will be done by using the technique developed in \cite{MCB} under the
condition
\begin{equation*}%\label{H-cond+}
  \lim_{\lambda\rightarrow0^+}\left[\mathrm{Tr}\, e^{-\lambda H_A}\right]^{\lambda}=1,
\end{equation*}
which is a slightly stronger than the Gibbs condition but holds for the Hamiltonians of many real quantum systems (see Section 2.2).

\section{Preliminaries}

\subsection{Basic notations}

Let $\mathcal{H}$ be a separable infinite-dimensional Hilbert space,
$\mathfrak{B}(\mathcal{H})$ the algebra of all bounded operators on $\mathcal{H}$ with the operator norm $\|\cdot\|$ and $\mathfrak{T}( \mathcal{H})$ the
Banach space of all trace-class
operators on $\mathcal{H}$  with the trace norm $\|\!\cdot\!\|_1$. Let
$\mathfrak{S}(\mathcal{H})$ be  the set of quantum states (positive operators
in $\mathfrak{T}(\mathcal{H})$ with unit trace) \cite{H-SCI,N&Ch,Wilde}.

Denote by $I_{\mathcal{H}}$ the identity operator on a Hilbert space
$\mathcal{H}$ and by $\id_{\mathcal{\H}}$ the identity
transformation of the Banach space $\mathfrak{T}(\mathcal{H})$.\smallskip

The \emph{von Neumann entropy} of a quantum state
$\rho \in \mathfrak{S}(\H)$ is  defined by the formula
$H(\rho)=\operatorname{Tr}\eta(\rho)$, where  $\eta(x)=-x\ln x$ for $x>0$
and $\eta(0)=0$. It is a concave lower semicontinuous function on the set~$\mathfrak{S}(\H)$ taking values in~$[0,+\infty]$ \cite{H-SCI,L-2,N&Ch,W,Wilde}.
The homogeneous concave extension of the von Neumann entropy to the cone~$\mathfrak{T}_+(\H)$ is given by the formula:
\begin{equation}\label{H-ext}
H(\rho)\doteq[\Tr\rho]H\biggl(\frac{\rho}{\Tr\rho}\biggr)=
\Tr\eta(\rho)-\eta(\Tr\rho).
\end{equation}
By using Theorem 11.10 in \cite{N&Ch} it
is easy to obtain the inequality
\begin{equation}\label{w-k-ineq+}
H\!\left(\sum_{k}p_k\rho_{k}\right)\leq
\sum_{k}p_kH(\rho_{k})+S\!\left(\{p_k\}\right),
\end{equation}
valid for any finite or countable collection  $\{\rho_{k}\}$ of positive operators in
the unit ball of $\mathfrak{T}(\mathcal{H})$ and any probability distribution $\{p_k\}$, where $S(\{p_k\})$ is the Shannon entropy.

In particular, for arbitrary positive operators $\rho$ and $\sigma$ in the unit ball of $\T(\H)$ and any $p\in(0,1)$
the following  inequality holds
\begin{equation}\label{w-k-ineq}
H(p\rho+(1-p)\sigma)\leq pH(\rho)+(1-p)H(\sigma)+h_2(p),
\end{equation}
where $\,h_2(p)=\eta(p)+\eta(1-p)\,$ is the binary entropy.\smallskip

By using inequality (\ref{w-k-ineq+}) it is easy to show that
\begin{equation}\label{w-k-ineq++}
H\!\left(\sum_{k}\rho_{k}\right)\leq
\sum_{k}H(\rho_{k})+S\!\left(\{\Tr\rho_{k}\}\right),
\end{equation}
for any finite or countable collection  $\{\rho_{k}\}$ of positive operators in
the unit ball of $\mathfrak{T}(\mathcal{H})$ such that $\sum_{k}\Tr\rho_{k}<+\infty$, where $S(\{x_k\})$ is the extended Shannon entropy
of the vector $\{x_k\}$ from the positive cone of $\ell_1$ defined as\footnote{It is easy to see that $S$ is the homogeneous extension of the "classical" Shannon entropy defined on the set of probability distributions to the positive cone of $\ell_1$.}
\begin{equation}\label{S-ext}
S\!\left(\{x_k\}\right)\doteq\sum_k\eta(x_k)-\eta\left(\sum_k x_k\right).
\end{equation}
Note that an equality holds in (\ref{w-k-ineq++}) if and only if the supports of all the operators $\rho_k$ are mutually orthogonal.\footnote{The support $\mathrm{supp}\rho$ of a positive trace class operator $\rho$ is the closed subspace spanned by the eigenvectors of $\rho$ corresponding to its positive eigenvalues.}

Uniform continuity bound for the von Neumann entropy in a finite dimensional quantum system was obtained by Fannes \cite{Fannes}. An optimized version of Fannes's continuity bound obtained by Audenaert in \cite{Aud} states that
\begin{equation}\label{Aud-CB}
 |H(\rho)-H(\sigma)|\leq \varepsilon \ln (d-1)+h_2(\varepsilon),\quad  \varepsilon=\textstyle\frac{1}{2}\|\rho-\sigma\|_1,
\end{equation}
for any states $\rho$ and $\sigma$ in a $d$-dimensional Hilbert space provided that $\varepsilon\leq  1-1/d$. \smallskip

\subsection{The set of quantum states with bounded energy}

One of the main results of this paper is the uniform continuity bound for the output entropy
of any positive linear map possessing property (\ref{PFE}) under the input energy constraint (Theorem \ref{OE-CB} in Section 5.2).
In this subsection we describe some preliminary results that are necessary for deriving this continuity bound.

Let $H_A$ be a positive (semi-definite) densely defined operator on a Hilbert space $\mathcal{H}_A$.  We will assume that
\begin{equation}\label{H-as}
\mathrm{Tr} H_A\rho=\sup_n\mathrm{Tr} P_n H_A\rho
\end{equation}
for any positive operator $\rho\in\T(\H_A)$, where $P_n$ is the spectral projector of $H_A$ corresponding to the interval $[0,n]$.\smallskip

Let $E_0$ be the infimum of the spectrum of $H_A$ and $E\geq E_0$. Then
\begin{equation}\label{s-b-e}
\mathfrak{C}_{H_A,E}=\left\{\rho\in\mathfrak{S}(\mathcal{H}_A)\,|\,\mathrm{Tr} H_A\rho\leq E\right\}
\end{equation}
is a closed convex subset of $\mathfrak{S}(\mathcal{H}_A)$. If
$H_A$ is treated as  Hamiltonian of a quantum system $A$ then
$\mathfrak{C}_{H_A,E}$  is the set of states with the mean energy not exceeding $E$.\smallskip

It is well known that the von Neumann entropy is continuous on the set $\mathfrak{C}_{H_A,E}$ for any $E> E_0$ if (and only if) the Hamiltonian  $H_A$ satisfies  the condition
\begin{equation}\label{H-cond}
  \mathrm{Tr}\, e^{-\lambda H_{A}}<+\infty\quad\textrm{for all}\;\lambda>0
\end{equation}
and that the maximal value of the entropy on this set is achieved at the \emph{Gibbs state} $\gamma_A(E)\doteq e^{-\lambda(E) H_A}/\mathrm{Tr} e^{-\lambda(E) H_A}$, where the parameter $\lambda(E)$ is determined by the equality $\mathrm{Tr} H_A e^{-\lambda(E) H_A}=E\mathrm{Tr} e^{-\lambda(E) H_A}$ \cite{W}. Condition (\ref{H-cond}) implies that $H_A$ is an unbounded operator having  discrete spectrum of finite multiplicity. It can be represented as follows
\begin{equation}\label{H-rep}
H_A=\sum_{k=0}^{+\infty} E_k |\tau_k\rangle\langle\tau_k|,
\end{equation}
where
$\left\{\tau_k\right\}_{k=0}^{+\infty}$ is the orthonormal
basis of eigenvectors of $H_A$ corresponding to the nondecreasing sequence $\left\{\smash{E_k}\right\}_{k=0}^{+\infty}$ of eigenvalues
tending to $+\infty$.

We will use the function
\begin{equation}\label{F-def}
F_{H_A}(E)\doteq\sup_{\rho\in\mathfrak{C}_{H_{\!A},E}}H(\rho)=H(\gamma_A(E)).
\end{equation}
It is easy to show that $F_{H_A}$ is a strictly increasing concave function on $[E_0,+\infty)$ such that $F_{H_A}(E_0)=\ln m(E_0)$, where $m(E_0)$ is the multiplicity of $E_0$ \cite{W-CB}.

In this paper we will assume that the Hamiltonian $H_A$ satisfies the condition
\begin{equation}\label{H-cond+}
  \lim_{\lambda\rightarrow0^+}\left[\mathrm{Tr}\, e^{-\lambda H_A}\right]^{\lambda}=1,
\end{equation}
which is slightly stronger than condition (\ref{H-cond}).\footnote{In terms of the sequence $\{E_k\}$ of eigenvalues of $H_A$
condition (\ref{H-cond}) means that $\lim_{k\rightarrow\infty}E_k/\ln k=+\infty$, while (\ref{H-cond+}) is valid  if $\;\liminf_{k\rightarrow\infty} E_k/\ln^q k>0\,$ for some $\,q>2$ \cite[Section 2.2]{MCB}.}
Condition (\ref{H-cond+}) holds if and only if
\begin{equation}\label{H-cond++}
  F_{H_A}(E)=o\shs(\sqrt{E})\quad\textrm{as}\quad E\rightarrow+\infty,
\end{equation}
while condition (\ref{H-cond}) is equivalent to  $\,F_{H_A}(E)=o\shs(E)\,$ as $\,E\rightarrow+\infty$ \cite[Section 2.2]{MCB}.
It is essential that condition (\ref{H-cond+})  holds for the Hamiltonians of many real quantum systems \cite{Datta}.\footnote{Theorem 3 in \cite{Datta} shows that $F_{H_A}(E)=O\shs(\ln E)$ as $E\rightarrow+\infty$ provided that condition (\ref{BD-cond}) holds.}

The function
\begin{equation}\label{F-bar}
 \bar{F}_{H_A}(E)=F_{H_A}(E+E_0)=H(\gamma_A(E+E_0))
\end{equation}
is concave and nondecreasing on $[0,+\infty)$. Let $\hat{F}_{H_A}$ be a continuous function
on $[0,+\infty)$ such that
\begin{equation}\label{F-cond-1}
\hat{F}_{H_A}(E)\geq \bar{F}_{H_A}(E)\quad \forall E>0,\quad \hat{F}_{H_A}(E)=o\shs(\sqrt{E})\quad\textrm{as}\quad E\rightarrow+\infty
\end{equation}
and
\begin{equation}\label{F-cond-2}
\hat{F}_{H_A}(E_1)<\hat{F}_{H_A}(E_2),\quad\hat{F}_{H_A}(E_1)/\sqrt{E_1}\geq \hat{F}_{H_A}(E_2)/\sqrt{E_2}
\end{equation}
for any $E_2>E_1>0$. Sometimes we will additionally assume that
\begin{equation}\label{F-cond-3}
\hat{F}_{H_A}(E)=\bar{F}_{H_A}(E)(1+o(1))\quad\textrm{as}\quad E\to+\infty.
\end{equation}
By property (\ref{H-cond++}) the role of $\hat{F}_{H_A}$ can be played by the function $\bar{F}_{H_A}$
provided that the function  $E\mapsto\bar{F}_{H_A}(E)/\sqrt{E}$ is nonincreasing. In general case the existence of a function $\hat{F}_{H_A}$ with the required properties is established in the following proposition proved in \cite{MCB}. \smallskip

\begin{property}\label{add-l}
A) \emph{If the Hamiltonian $H_A$ satisfies condition (\ref{H-cond+}) then
\begin{equation}\label{m-fun}
\hat{F}^{*}_{H_A}(E)\doteq \sqrt{E}\sup_{E'\geq E}\bar{F}_{H_A}(E')/\sqrt{E'}
\end{equation}
is the minimal function satisfying all the conditions in (\ref{F-cond-1}) and (\ref{F-cond-2}).}

B) \emph{Let
\begin{equation*}
\!N_{\shs\uparrow}[H_A](E)\doteq \sum_{k,j: E_k+E_j\leq E} E_k^2\quad \textrm{and}\quad N_{\downarrow}[H_A](E)\doteq \sum_{k,j: E_k+E_j\leq E} E_kE_j
\end{equation*}
for any $E>E_0$. If
\begin{equation}\label{BD-cond}
\exists \lim_{E\rightarrow+\infty}N_{\shs\uparrow}[H_A](E)/N_{\downarrow}[H_A](E)=a>1
\end{equation}
then
\begin{itemize}
  \item  there is $E_*$ such that the function $E\mapsto\bar{F}_{H_A}(E)/\sqrt{E}$ is nonincreasing for all $E\geq E_*$ and hence $\hat{F}^{*}_{H_A}(E)=\bar{F}_{H_A}(E)$ for all $E\geq E_*$;
  \item $\hat{F}^{*}_{H_A}(E)=(a-1)^{-1} (\ln E)(1+o(1))$ as $E\rightarrow+\infty$.
\end{itemize}}
\end{property}
\medskip

In \cite{Datta} it is shown that condition (\ref{BD-cond}) is valid for the Hamiltonians
of many real quantum systems.\smallskip

Practically, it is convenient to use functions $\hat{F}_{H_A}$ defined by simple formulae. The example of
such function $\hat{F}_{H_A}$ satisfying all the conditions in (\ref{F-cond-1}),(\ref{F-cond-2}) and (\ref{F-cond-3}) in the case when $A$ is a multimode quantum oscillator is considered in Section 5.2.\smallskip

\section{On positive linear maps  preserving finiteness of the entropy}

Many results concerning quantum channels preserving finiteness of the entropy are valid
for arbitrary positive linear maps possessing this property, i.e. property (\ref{PFE}), where
the entropy $H(\Phi(\rho))$ of a positive operator $\Phi(\rho)$ is defined according to formula (\ref{H-ext}).

We will call such maps (channels, operations) PFE-maps (cnannels, operations) for brevity.\footnote{In \cite{OE,PCE} these maps were
called PCE-maps, since they also preserve local continuity of the entropy by Theorem \ref{main} below.}

\subsection{Characterization of PFE-maps}

The following theorem proved in \cite{PCE} gives characterisations of the class (semigroup) of
all PFE-maps. \smallskip

\begin{theorem}\label{main}
\emph{Let $\,\Phi$~be a  positive linear map from~$\mathfrak{T}(\H_A)$ into~$\mathfrak{T}(\H_B)$.
The following properties are equivalent:}\smallskip

{\rm(i)} \emph{$\,\Phi$~preserves finiteness of the entropy, i.e. property (\ref{PFE}) holds};
\smallskip

{\rm(ii)} \emph{$\,\Phi$~preserves continuity of the entropy, i.e. property (\ref{PCE}) holds};
\smallskip

{\rm(iii)} \emph{the output entropy ~$H_{\Phi}$ is bounded on the set $\,\mathrm{ext}\mathfrak{S}(\H_A)$ of pure states.}\footnote{Pure states are one-rank projectors -- extreme points of the convex set $\mathfrak{S}(\H_A)$.}\medskip
\end{theorem}

Theorem \ref{main}  implies, in particular, that  property (\ref{PFE}) holds if and only if
\begin{equation}\label{PFE+}
H^\mathrm{p}_{\max}(\Phi)\doteq\sup_{\rho\in\mathrm{ext}\S(\H_A)}H_{\Phi}(\rho)<+\infty.
\end{equation}

The parameter $H^\mathrm{p}_{\max}(\Phi)$ will be used below in quantitative continuity analysis of the
function $H_{\Phi}$. It can be estimated by using a concrete expression for the map $\Phi$. Below, in Section 3.2,
we obtain an upper bound on $H^\mathrm{p}_{\max}(\Phi)$ by using the Kraus representation for $\Phi$.\smallskip

By  using inequality (\ref{w-k-ineq+}) and the spectral decomposition of an arbitrary  state $\rho$ in $\S(\H_A)$  it is easy to show that
\begin{equation}\label{H-ub}
H_{\Phi}(\rho)\leq H(\rho)+H^\mathrm{p}_{\max}(\Phi).
\end{equation}

The simplest  PFE-maps are maps with finite-dimensional output space and unitary transformations, i.e. maps of the form $\Phi(\rho)=U\rho U^*$, where  $U$~is an isometry from~$\H_A$ into~$\H_B$. More interesting examples are considered in the next subsections.

\subsection{Completely positive maps}

In this subsection we apply the criteria in Theorem \ref{main} to the class of quantum channels and operations
--  completely positive trace-preserving and trace-non-increasing linear maps \cite{H-SCI,N&Ch,Wilde}.

It is well known that any quantum operation (correspondingly, channel) $\Phi$ has the Kraus representation
\begin{equation}\label{Kraus}
\Phi(\rho)=\sum_{k}V_k\rho V_k^*,
\end{equation}
where $\{V_k\}$ is a collection of linear operators from $\H_A$ to $\H_B$ such that $\sum_{k}V_k^*V_k\leq I_{\H_A}$ (correspondingly, $\sum_{k}V_k^*V_k=I_{\H_A}$).
The minimal number of nonzero summands in representation (\ref{Kraus}) is called \emph{Choi rank} of the operation $\Phi$ \cite{H-SCI,N&Ch,Wilde}.

If a quantum operation $\Phi$ has finite Choi rank $m$ then it follows from (\ref{H-ub})  that $H_{\Phi}(\rho)\leq \ln m$ for any pure state $\rho$. Hence, Theorem \ref{main} shows that $\Phi$ is a PFE\nobreakdash-\hspace{0pt}operation.\smallskip

Inequality (\ref{w-k-ineq++}) implies that for any unit vector $\varphi$ in $\H_A$ the following inequality holds
\begin{equation}\label{u-in}
 H(\Phi(|\varphi\rangle\langle\varphi|))=H\!\left(\sum_{k}V_k |\varphi\rangle\langle\varphi|V_k^*\right)\leq S\!\left(\{\|V_k\varphi\|^2\}_k\right),
\end{equation}
where $S(\cdot)$ is the extended Shannon entropy defined in (\ref{S-ext}). Inequality (\ref{u-in}) shows that
\begin{equation}\label{u-in+}
H^\mathrm{p}_{\max}(\Phi)\leq \sup_{\varphi\in\H^1_A}S\!\left(\{\|V_k\varphi\|^2\}_k\right)\leq S\!\left(\left\{\|V_{k}\|^{2}\right\}_{k}\right),
\end{equation}
where $\H_A^1$ is the unit sphere of $\H_A$ and it is assumed that $\,S\!\left(\left\{\|V_{k}\|^{2}\right\}_{k}\right)=+\infty$
if $\,\sum_k\!\|V_{k}\|^{2}=+\infty$.\smallskip

Theorem \ref{main} implies the following conditions of the PFE\nobreakdash-\hspace{0pt}property (\ref{PFE}) for quantum operations with infinite Choi rank.\smallskip

\begin{corollary}\label{main-c3} \emph{A quantum operation  $\Phi$ having representation (\ref{Kraus}) possesses the PFE\nobreakdash-\hspace{0pt}property (\ref{PFE}) if one of the following conditions holds:}\medskip

{\rm a)} \emph{the function $\,\varphi\mapsto S\!\left(\{\|V_k\varphi\|^2\}_k\right)\,$ is bounded on the unit sphere of $\,\H_A$;}\medskip

{\rm b)} \emph{$\sum_k\!\|V_{k}\|^{2}$ and  $\,S\!\left(\left\{\|V_{k}\|^{2}\right\}_{k}\right)$ are finite;}\medskip

{\rm c)}  \emph{there exists a sequence $\{h_{k}\}$ of nonnegative numbers
such that
$$
\left\|\sum_{k}h_{k}V^{*}_{k}V_{k}\right\|<+\infty\quad\textit{and}\quad
\sum_{k}e^{-h_{k}}<+\infty.
$$}

\textit{If $\;\mathrm{Ran}V_{k}\perp\mathrm{Ran}V_{j}$ for all
$\,k\neq j$ then $\,\mathrm{a)}$ is a
necessary and sufficient condition of property (\ref{PFE}) for the
operation $\,\Phi$.}
\end{corollary}\medskip

\emph{Proof.} By Theorem \ref{main} the sufficiency of conditions a) and b)  follows from inequality (\ref{u-in+}).
The necessity of condition a) in the case $\;\mathrm{Ran}V_{k}\perp\mathrm{Ran}V_{j}$ follows from the remark after inequality (\ref{w-k-ineq++}).

To prove
the implication $\,\mathrm{c)\Rightarrow a)}\,$  it suffices to note that the extended Shannon entropy is bounded on the subset of the positive part of the unit ball in $\ell_1$ consisting of vectors $\{p_k\}$ such that $\sum_k h_kp_k\leq C$ for any $C$. $\square$\medskip

\begin{remark}\label{main-c3-r} Condition b) in Corollary \ref{main-c3} is the most easily verified but is too rough  because it depends only on the norms of the Kraus operators. Condition c) is more subtle, since it takes  "geometry" of the sequence $\{V_k\}$ into account. This is confirmed by the following example.
\end{remark}\smallskip

\begin{example}\label{exam-1} Let $\{P_k\}_{k> 1}$ be any sequence of mutually orthogonal projectors in $\B(\H_A)$ and $\alpha\in[0,\ln 2]$.
Consider the quantum channel
\begin{equation*}%\label{Kraus}
\Phi_{\alpha}(\rho)=\sum_{k\geq 1}c_k P_k\rho P_k,
\end{equation*}
where $c_1=1$, $\,c_k=\alpha/\ln k\;$ for $\,k>1\,$ and $\,P_1=\sqrt{I_{\H_A}-\sum_{k>1}c_k P_k}$. Condition c) in Corollary \ref{main-c3} shows that $\Phi_\alpha$ is a PFE\nobreakdash-\hspace{0pt}channel, while condition b) is not valid in this case.
\end{example}

Sometimes it is possible to prove PFE-property (\ref{PFE}) of a quantum channel (operation) without using its Kraus representation.\smallskip

\begin{example}\label{exam-2} Let $\mathcal{H}_{a}$ be the Hilbert space
$\mathcal{L}_{2}([-a,+a])$, where $a<+\infty$, and
$\{U_{t}\}_{t\in\mathbb{R}}$ be the group of unitary operators in
$\mathcal{H}_{a}$ defined as follows
$$
(U_{t}\varphi)(x)=e^{-\mathrm{i}tx}\varphi(x),\quad\forall\varphi\in\mathcal{H}_{a}.
$$
For given probability density function $p(t)$ consider the quantum
channel
$$
\Phi_{p}^{a}:\mathfrak{T}(\mathcal{H}_{a})\ni
\rho\mapsto\int_{-\infty}^{+\infty}U_{t}\rho
U_{t}^{*}p(t)dt\in\mathfrak{T}(\mathcal{H}_{a}).
$$
One can show that the function $H_{\Phi_{p}^{a}}$ is  bounded (and continuous) on the set
$\mathrm{ext}\S(\H_a)$ provided that the
differential entropy of the distribution $p(t)$ is finite and that
the function $p(t)$ is bounded and monotonic on $(-\infty,-b]$ and
on $[+b,+\infty)$ for sufficiently large $b$ \cite[Example 3]{OE}. So, in this case
$\Phi_{p}^{a}$ is a PFE\nobreakdash-\hspace{0pt}channel with infinite Choi rank.
\end{example}

\subsection{Types of PFE-channels and tensor products}

It follows from the definition that the class of PFE\nobreakdash-\hspace{0pt}channels is closed under composition: if $\Phi:A\rightarrow B$ and $\Psi:B\rightarrow C$ are PFE\nobreakdash-\hspace{0pt}channels then $\Psi\circ\Phi$ is a PFE\nobreakdash-\hspace{0pt}channel. Inequality (\ref{w-k-ineq}) shows that any convex mixture of
PFE\nobreakdash-\hspace{0pt}channels between given quantum systems is a PFE\nobreakdash-\hspace{0pt}channel. But the class of PFE\nobreakdash-\hspace{0pt}channels is not closed under tensor products: the tensor product of the identity channel and the completely depolarising channel with a pure output state is not a PFE\nobreakdash-\hspace{0pt}channel. Moreover, it is easy to see that the tensor square of the PFE\nobreakdash-\hspace{0pt}channel $\rho\mapsto (1-p)\rho+p\shs\sigma$, where $\sigma$ is a given pure state, is not a PFE\nobreakdash-\hspace{0pt}channel. On the other hand, there are nontrivial
PFE\nobreakdash-\hspace{0pt}channels $\Phi$ and $\Psi$ such that $\Phi\otimes\Psi$ is a PFE\nobreakdash-\hspace{0pt}channel (see below). Note also that the
PFE\nobreakdash-\hspace{0pt}property (\ref{PFE}) of a channel $\Phi\otimes\Psi$ obviously implies the same property of the channels $\Phi$ and $\Psi$.

To analyse the PFE\nobreakdash-\hspace{0pt}property (\ref{PFE}) of tensor products we will introduce a classification of
PFE\nobreakdash-\hspace{0pt}channels based on the notion of a complementary channel.\smallskip

For any quantum channel $\,\Phi:A\rightarrow B\,$ the Stinespring theorem implies existence of a Hilbert space
$\mathcal{H}_E$ and of an isometry
$V:\mathcal{H}_A\rightarrow\mathcal{H}_B\otimes\mathcal{H}_E$ such
that
\begin{equation*}%\label{St-rep}
\Phi(\rho)=\mathrm{Tr}_{E}V\rho V^{*},\quad
\rho\in\mathfrak{T}(\mathcal{H}_A).
\end{equation*}
The quantum  channel
\begin{equation*}%\label{c-channel}
\mathfrak{T}(\mathcal{H}_A)\ni
\rho\mapsto\widehat{\Phi}(\rho)=\mathrm{Tr}_{B}V\rho
V^{*}\in\mathfrak{T}(\mathcal{H}_E)
\end{equation*}
is called \emph{complementary} to the channel $\Phi$
\cite[Ch.6]{H-SCI}.\smallskip

Since the functions $H_{\Phi}$ and $H_{\widehat{\Phi}}$ coincide on the sets of pure states in $\S(\H_A)$, Theorem \ref{main} implies the following\smallskip
\begin{corollary}\label{comp-ch}
\emph{$\,\Phi$ is a PFE-channel if and only if $\,\widehat{\Phi}$ is a PFE-channel.}
\end{corollary}
\smallskip

The function $H_{\widehat{\Phi}}$ is an important entropic characteristic of a channel $\Phi$ called the entropy exchange of $\Phi$ \cite{H-SCI,Wilde}.
\smallskip

\begin{property}\label{PCE-types} \emph{Let A be an infinite-dimensional quantum system. }
\emph{Any PFE-chan- nel $\,\Phi:A\rightarrow B$ belongs to one of the classes $\mathcal{A}$,$\,\mathcal{B}$ and $\,\mathcal{C}$ characterized, respectively, by the conditions:}
\begin{enumerate}[a)]
  \item \emph{$H_{\Phi}(\rho)<+\infty$ for any state $\,\rho\in\S(\H_A)$;}
  \item \emph{$H_{\widehat{\Phi}}(\rho)<+\infty$ for any state $\,\rho\in\S(\H_A)$;}
  \item \emph{$\!\sup\limits_{\rho\in\mathrm{ext}\mathfrak{S}(\H_A)}\!\!\!H_{\Phi}(\rho)<\!+\infty$, but $H_{\Phi}(\rho)=\!H_{\widehat{\Phi}}(\rho)=\!+\infty$ for some state $\rho\in\S(\H_A)$.}
\end{enumerate}
\emph{The classes $\mathcal{A}$,$\,\mathcal{B}$ and $\,\mathcal{C}$ are convex. Convex mixtures of channels from different classes belong to the class $\mathcal{C}$.}\medskip

\emph{The classes $\mathcal{A}$ and $\mathcal{B}$ are closed under composition\footnote{in the sense described at the begin of this subsection}, while the class $\mathfrak{C}$ is not.}

\end{property}

\medskip

\emph{Proof.} By Theorem 1 in \cite{OE} condition a) (correspondingly, b)) implies boundedness and continuity of the function
$H_{\Phi}$ (correspondingly, $H_{\widehat{\Phi}}$) on the set $\S(\H_A)$. So, any of these conditions implies the PFE\nobreakdash-\hspace{0pt}property (\ref{PFE}) of the channel $\Phi$. The inequality
$$
H(\rho)\leq H_{\Phi}(\rho)+H_{\widehat{\Phi}}(\rho),\quad \rho\in \S(\H_A)
$$
(which follows from subadditivity of the entropy) shows that the classes $\A$
and $\B$ are disjoint. The convexity of all the classes can be established by using basic properties of the entropy and the relation
\begin{equation*}%\label{ee-rel}
H_{\widehat{\Phi}}(\rho)=H_{\Phi\otimes\id_R}(\hat{\rho}),\quad \rho\in \S(\H_A),
\end{equation*}
where $\hat{\rho}$ is a purification of $\rho$ in $\S(\H_{AR})$ \cite{H-SCI}.

The closedness of the class $\A$ under composition is obvious.
To prove the same property of the class $\B$ we will use Proposition \ref{TP} below (proved independently).

Assume that  $\Phi:A\rightarrow B$ and $\Psi:B\rightarrow C$ are PFE\nobreakdash-\hspace{0pt}channels of the class $\B$. Let $R$ be an infinite-dimensional quantum system. By Proposition \ref{TP}A $\Phi\otimes\id_R$ and $\Psi\otimes\id_R$ are PFE\nobreakdash-\hspace{0pt}channels. So,
$[\Psi\otimes\id_R]\circ[\Phi\otimes\id_R] =[\Psi\circ\Phi]\otimes\id_R$ is a PFE\nobreakdash-\hspace{0pt}channel. By Proposition \ref{TP}B
$\Psi\circ\Phi$ is a PFE\nobreakdash-\hspace{0pt}channel from the class $\B$.

To show that the class $\C$ is not closed under composition consider the PFE-channels
$$
\Phi(\rho)=P\rho P\oplus [\Tr\bar{P}\rho]\sigma\quad\textrm{ and }\quad \Psi(\rho)=[\Tr P\rho]\varsigma\oplus \bar{P}\rho \bar{P},
$$
where $P$ and $\bar{P}=I_{\H}-P$ are infinite rank projectors, $\sigma$ and $\varsigma$ are pure states such that $\bar{P}\sigma\bar{P}=\sigma$ and $P\varsigma P=\varsigma$. These channels belong to the class $\C$. The simplest way to show this is to note that $\Phi^{\otimes2}$ and $\Psi^{\otimes2}$ are not PFE-channels and to use Proposition \ref{TP}A below. It is easy to see that $\Psi\circ\Phi$ is a completely depolarizing channel belonging to the class $\A$. $\square$ \smallskip

The class $\A$ consists of channels with continuous output entropy. The simplest example of such channels is the completely depolarizing channel $\rho\mapsto [\Tr\rho]\sigma$, where $\sigma$ is a given state with finite entropy. More interesting channels from the class $\A$ are considered in \cite[Section 3]{OE}.\smallskip

The class $\B$ contains  the identity channel and all channels with finite Choi rank. The channel from the class $\B$ having infinite Choi rank is presented in the above Example \ref{exam-2}. The finiteness of the function $H_{\widehat{\Phi}_{p}^{a}}$ in this case can be shown by using the explicit expression for the channel $\widehat{\Phi}_{p}^{a}$ and the results in \cite[Section 3]{OE}.\smallskip

The simplest example of a PFE-channel from the class $\C$ is the quantum channel $\rho\mapsto (1-p)\rho+p\sigma$, where $\sigma$ is a pure state, -- the convex mixture of the identity channel (from the class $\B$) and the completely depolarising channel (from the class $\A$).\smallskip

\begin{property}\label{TP} \emph{Let $\,\Phi$ and $\,\Psi$ be PFE-channels with infinite-dimensional input systems.}%\smallskip

A) \emph{If both channels $\,\Phi$ and $\,\Psi$ belong to one of the classes $\A$ and $\B$ then $\Phi\otimes\Psi$ is a PFE-channel of the same class.} \smallskip

B) \emph{If the channels $\,\Phi$ and $\,\Psi$ belong to different classes then $\,\Phi\otimes\Psi$ is not a PFE-channel.}\smallskip
\end{property}

\emph{Proof.} A) If $\Phi$ and $\Psi$ are channels of the class $\A$ (correspondingly, the class $\B$) then the functions $H_{\Phi}$ and $H_{\Psi}$ (correspondingly, the functions $H_{\widehat{\Phi}}$ and $H_{\widehat{\Psi}}$) are bounded. By subadditivity of the entropy this implies boundedness of the function $H_{\Phi\otimes\Psi}$
(correspondingly, the function $H_{\widehat{\Phi}\otimes\widehat{\Psi}}=H_{\widehat{\Phi\otimes\Psi}}$).
\smallskip

B) Let $\Phi:A\rightarrow B$ be a channel of the class $\A$ and $\Psi:C\rightarrow D$ a channel of one of the classes $\B$ and $\C$. Let $\rho$ be a state in $\S(\H_C)$ such that $H_{\Psi}(\rho)=+\infty$ and $\hat{\rho}$ a purification of $\rho$ in $\S(\H_{AC})$. Since $H_{\Phi}(\hat{\rho}_A)<+\infty$, the triangle inequality
$$
H_{\Phi\otimes\Psi}(\hat{\rho})\geq \left|H_{\Phi}(\hat{\rho}_A)-H_{\Psi}(\hat{\rho}_C)\right|=\left|H_{\Phi}(\hat{\rho}_A)-H_{\Psi}(\rho)\right|
$$
shows that $H_{\Phi\otimes\Psi}(\hat{\rho})=+\infty$.

If channels $\Phi$  and $\Psi$ belong, respectively, to the classes $\B$ and $\C$ then the similar arguments applied to the complementary channels
$\widehat{\Phi}$  and $\widehat{\Psi}$  show that $\widehat{\Phi\otimes\Psi}$ is not a PFE-channel. This means, by Corollary \ref{comp-ch}, that
$\Phi\otimes\Psi$ is not a PFE-channel. $\square$\medskip

\begin{remark}\label{TP-r} Proposition \ref{TP} says nothing about tensor product of channels from the class $\C$. There is a conjecture that $\Phi\otimes\Psi$ is not a PFE-channel if one of the channels, say $\Phi$, belongs to the class $\C$. If $\Phi$ is a convex mixture of channels
from the classes $\A$ and $\B$ then this assertion directly follows from Proposition \ref{TP}B, but it is not clear how to prove it in general case.
\end{remark}\medskip

\begin{corollary}\label{TP-c} \emph{If $\,\Phi$ is a  PFE-channel from  one of the classes $\A$ and $\B$ then $\,\Phi^{\otimes n}$ is a PFE-channel of the same class for any $\,n$.}
\end{corollary}

\section{The convex closure of the output entropy}

\subsection{General results}

In analysis of informational properties of a quantum channel $\Phi$ the convex closure $\,\overline{\mathrm{co}}H_{\Phi}$ of its output entropy plays important role \cite{H-SCI,EM}. The function $\,\overline{\mathrm{co}}H_{\Phi}$ is defined as the maximal closed (lower semicontinuous) convex function on $\S(\H_A)$ majorized by the function $H_{\Phi}$. In finite dimensions $\,\overline{\mathrm{co}}H_{\Phi}$ coincides with the convex hull $\,\mathrm{co}H_{\Phi}$ of $H_{\Phi}$ -- the maximal convex function on $\S(\H_A)$ majorized by the function $H_{\Phi}$ which is given by the formula
\begin{equation}\label{c-hull}
  \mathrm{co}H_{\Phi}(\rho)=\inf_{\sum_ip_i\rho_i=\rho}\sum_i p_iH_{\Phi}(\rho_i),
\end{equation}
where the infimum is over all finite ensembles $\{p_i, \rho_i\}$ of input states with the average state $\rho$.

In infinite dimensions the function $\,\overline{\mathrm{co}}H_{\Phi}$ coincides with $\,\mathrm{co}H_{\Phi}$
only for positive maps (channels) with finite output entropy, but one can assume that it coincides with the $\sigma$-convex hull $\sigma\textrm{-}\mathrm{co}H_{\Phi}$ of $H_{\Phi}$ defined by formula (\ref{c-hull}) in which the infimum is over all countable ensembles $\{p_i, \rho_i\}$ of input states with the average state $\rho$.

On the other hand, the compactness criterion for families  of probability measures on $\S(\H)$
makes it possible to show that
\begin{equation}\label{c-clos}
\overline{\mathrm{co}}H_{\Phi}(\rho)=\inf_{\bar{\rho}(\mu)=\rho}\int H_{\Phi}(\rho)\mu(d\rho),
\end{equation}
where the infimum is over all Borel probability measures on the set $\S(\H_{A})$  with the barycenter $\rho$ \cite{EM}.
So, to prove the conjecture $\sigma\textrm{-}\mathrm{co}H_{\Phi}=\overline{\mathrm{co}}H_{\Phi}$ it suffices to show that the infimum in
(\ref{c-clos}) can be taken only over all discrete probability measures. We don't know how to prove (or disprove) this conjecture in general\footnote{It is shown in \cite{EM} that $\,\sigma\textrm{-}\mathrm{co}f\neq\overline{\mathrm{co}}f\,$ for a particular  lower semicontinuous concave nonnegative unitarily invariant function $f$ on $\S(\H)$, so the above conjecture can not be proved by using only general entropy-type properties of the function $H_{\Phi}$.}, but it seems reasonable to mention that it holds for PFE\nobreakdash-\hspace{0pt}channels.\smallskip

\begin{corollary}\label{main-c2} \emph{Let $\Phi:\T(\H_A)\rightarrow\T(\H_B)$  be a positive map possessing PFE-property (\ref{PFE}). Then}\medskip

{\rm A)} \emph{$\sigma\textup{-}\mathrm{co}H_{\Phi}(\rho)=\overline{\mathrm{co}}H_{\Phi}(\rho)$ for any state $\rho\in\mathfrak{S}(\H_A)$;}\medskip

{\rm B)} \emph{the function~$\,\sigma\textup{-}\mathrm{co}H_{\Phi}=\overline{\mathrm{co}}H_{\Phi}\,$ is continuous and bounded on
$\,\mathfrak{S}(\H_A)$.}
\end{corollary}\smallskip

\emph{Proof.} Since the assumption implies,  by Theorem \ref{main}, continuity and boundedness of the function $H_{\Phi}$ on the set $\mathrm{ext}\S(\H_A)$, both assertions of the corollary can be  derived from Corollary 2 in \cite{EM} by using Remark \ref{main-r} below.\smallskip

\begin{remark}\label{main-r} By using concavity and lower semicontinuity of the function $H_{\Phi}$ one can show that $\,\sigma\textrm{-}\mathrm{co}H_{\Phi}=\check{H}_{\Phi}^{\sigma}\,$ and $\,\overline{\mathrm{co}}H_{\Phi}=\check{H}_{\Phi}^{\mu}\,$, where $\check{H}_{\Phi}^{\sigma}$ and $\check{H}_{\Phi}^{\mu}$ are discrete and continuous convex roof extensions\footnote{The convex roof extension is widely used for construction of different characteristics of states in finite-dimensional quantum systems \cite{4H,P&V}.} of the function $H_{\Phi}|_{\mathrm{ext}\S(\H_A)}$  defined, respectively, by the right hand sides of (\ref{c-hull}) and  (\ref{c-clos}) in which the infima are over all ensembles (measures) consisting of (supported by) pure states \cite[Section 2.3]{EM}. Thus, the assertion of Corollary \ref{main-c2} can be reformulated in terms of the functions  $\check{H}_{\Phi}^{\sigma}$ and $\check{H}_{\Phi}^{\mu}$ (instead of $\sigma\textrm{-}\mathrm{co}H_{\Phi}$ and $\overline{\mathrm{co}}H_{\Phi}$).
\end{remark}

\subsection{Applications to the entanglement theory}

Important task of the entanglement theory consists in finding appropriate characteristics of entanglement of  composite states and in  exploring their properties \cite{Eisert,4H}.

If $\rho_{AB}$ is a pure state of a bipartite system $AB$ of any dimension then its entanglement is characterized by the von Neumann entropy of partial states:
$$
E(\rho_{AB})\doteq H(\rho_A)=H(\rho_B).
$$

Entanglement of mixed states of a bipartite system $AB$ is characterized by different entanglement measures \cite{ESP,P&V,V}. One of the most important of them is the Entanglement of Formation (EoF).

In the case of finite-dimensional bipartite system $AB$ the EoF is defined as the convex roof extension to the set $\S(\H_{AB})$ of the function $\rho_{AB}\mapsto H(\rho_{A})$ on the set $\mathrm{ext}\shs\S(\H_{AB})$ of pure states, i.e.
\begin{equation}\label{ef-def}
  E_{F}(\rho_{AB})=\inf_{\sum_i p_i\rho_{AB}^i=\rho_{AB}}\sum_i p_iH(\rho^i_{A}),
\end{equation}
where the infimum is over all ensembles $\{p_i, \rho_{AB}^i\}$ of pure states with the average state $\rho_{AB}$ \cite{Bennett}.\smallskip

In infinite dimensions there are two versions $E_F^{d}$ and $E_F^{c}$ of the EoF defined, respectively, by using discrete and continuous convex roof extensions, i.e.
$$
E_F^{d}(\rho_{AB})=\!\inf_{\sum_i\!p_i\rho^i_{AB}=\rho_{AB}}\sum_ip_iH(\rho^i_A),\;\;\;\;\; E_F^{c}(\rho_{AB})=\!\inf_{b(\mu)=\rho_{AB}}\int\! H(\varrho_A)\mu(d\varrho_{AB}),
$$
where the first infimum is over all countable convex decompositions of the state $\rho_{AB}$ into pure states and the second one is over all Borel probability measures on the set $\mathrm{ext}\shs\S(\H_{AB})$ with the barycenter $\rho_{AB}$ \cite[Section 5]{EM}.

The continuous version $E_F^{c}$ is a lower semicontinuous function on the set $\S(\H_{AB})$ of all states of infinite-dimensional bipartite system possessing basic properties of entanglement measures (including monotonicity under generalized selective measurements) \cite{EM}. The discrete version $E_F^{d}$ seems more preferable from the physical point of view but the assumption $E_F^{d}\neq E_F^{c}$ leads to several problems with this version, in particular, it is not clear how to prove its vanishing for  countably nondecomposable separable states.\footnote{In general, the discrete convex roof construction applied to an entropy type function (in the role of $H$) may give a function which is not equal to zero at  countably non-decomposable separable states \cite[Remark 6]{EM}.}

In \cite{EM} it is shown that $E_F^{d}(\rho_{AB})=E_F^{c}(\rho_{AB})$ for any state $\rho_{AB}$ such that $$
\min\{H(\rho_{A}),H(\rho_{B}),H(\rho_{AB})\}<+\infty,$$
but the coincidence of $E_F^{d}$ and $E_F^{c}$ on the whole set $\S(\H_{AB})$ is not proved yet (as far as we know).
It is equivalent to the lower semicontinuity of $E_F^{d}$ on $\S(\H_{AB})$ (since $E_F^{c}$ coincides with the convex closure of the entropy of a partial trace).

By applying Corollary \ref{main-c2} and Remark \ref{main-r} to the channel $\,\Phi(\rho_{AB})=\rho_{A}$ we obtain  the following
\smallskip

\begin{property}\label{Ef} \emph{Let $\,\K$ be a subspace of $\,\H_{AB}$ such that all unit vectors in $\,\K$ have  bounded entanglement, i.e.
$\,\sup_{\varphi\in\K, \|\varphi\|=1}E(|\varphi\rangle\!\langle\varphi|)<+\infty$. Then}

 \smallskip

\rm A) \emph{$E^c_F(\rho)=E^d_F(\rho)$ for any state $\rho$ in $\,\S(\K)$;}\smallskip

\rm B) \emph{the function $E^c_F=E^d_F$ is continuous on the set $\,\S(\K)$.}
\end{property}\smallskip

The existence of nontrivial subspaces satisfying the condition of Proposition \ref{Ef} follows, by the Stinespring representation, from the existence of PFE\nobreakdash-\hspace{0pt}channels of the class $\B$ with infinite Choi rank and of PFE\nobreakdash-\hspace{0pt}channels of the class $\C$ (see Section 3).

\section{Uniform continuity bounds for the output entropy}

\subsection{Positive linear maps  with finite output entropy}

Assume that $\Phi$ is a positive trace-non-increasing linear map from $\T(\H_A)$ to $\T(\H_B)$
such that
\begin{equation}\label{FOE}
H_{\Phi}(\rho)\doteq H(\Phi(\rho))<+\infty\;\textrm{ for any state}\;\rho\;\textrm{ in  }\;\S(\H_A).
\end{equation}
The simplest example of such a map is a map with finite-dimensional output system, i.e. when $\dim\H_B<+\infty$. Nontrivial examples
of quantum channels satisfying condition (\ref{FOE}) are considered in \cite{OE}.

By Theorem 1 in \cite{OE} for any positive trace-non-increasing linear map $\Phi$ possessing property (\ref{FOE})
the function $H_{\Phi}$ is continuous and bounded on the whole set $\S(\H_A)$ of input states. The concavity of the entropy and inequality
(\ref{w-k-ineq}) imply that
\begin{equation}\label{LAA}
0\leq H_{\Phi}(p\rho+(1-p)\sigma)-\left(pH_{\Phi}(\rho)+(1-p)H_{\Phi}(\sigma)\right)\leq h_2(p),\quad p\in[0,1],
\end{equation}
for any states $\rho$ and $\sigma$ in $\S(\H_A)$, where $\,h_2(p)\,$ is the binary entropy.\smallskip

Note also that for any states $\rho$ and $\sigma$ in $\S(\H_A)$ we have
\begin{equation}\label{w-k-ineq+++}
|H_{\Phi}(\rho)-H_{\Phi}(\sigma)|\leq H_{\max}(\Phi)-H_{\min}(\Phi),
\end{equation}
where
$$
H_{\min}(\Phi)=\inf_{\rho\in\S(\H_A)}H_{\Phi}(\rho)\geq0 \quad\textrm{ and }\quad H_{\max}(\Phi)=\sup_{\rho\in\S(\H_A)}H_{\Phi}(\rho)<+\infty.
$$

Inequalities (\ref{LAA}) and (\ref{w-k-ineq+++}) allow to apply the Alicki-Fannes-Winter method (proposed in the optimal form in \cite{W-CB} and
described in a full generality in the proof of Proposition 1 in \cite{CMI}) to the function $H_{\Phi}(\rho)$. As a result we obtain\smallskip

\begin{property}\label{s-c-b} \emph{Let $\Phi:\T(\H_A)\rightarrow\T(\H_B)$  be a positive trace-non-increasing linear map possessing property (\ref{FOE}). Then
\begin{equation}\label{OE-CB-1}
  |H_{\Phi}(\rho)-H_{\Phi}(\sigma)|\leq \varepsilon [H_{\max}(\Phi)-H_{\min}(\Phi)]+g(\varepsilon)
\end{equation}
for any states $\rho$ and $\sigma$ in $\,\S(\H_A)$, where $\;\varepsilon=\frac{1}{2}\|\shs\rho-\sigma\|_1$ and $\,g(\varepsilon)=(1+\varepsilon)h_2\left(\frac{\varepsilon}{1+\varepsilon}\right)$.}
\end{property}\smallskip

\begin{remark}\label{s-c-br-2} The l.h.s. of (\ref{OE-CB-1}) can be made arbitrarily close to $H_{\max}(\Phi)-H_{\min}(\Phi)$
by appropriate choice of $\rho$ and $\sigma$. Since $\frac{1}{2}\|\shs\rho-\sigma\|_1\leq 1$  for any $\rho$ and $\sigma$, this shows that
continuity bound (\ref{OE-CB-1}) is close to tight when $H_{\max}(\Phi)-H_{\min}(\Phi)\gg g(1)=2\ln2$.
\end{remark}\smallskip

If $d_B\doteq\dim\H_B<+\infty$ then $H_{\max}(\Phi)$ in (\ref{OE-CB-1}) can be replaced by $k\log d_B$, where $k=\sup_{\rho\in\S(\H_A)}\Tr \Phi(\rho)$.
In this case  continuity bound for the function $H_{\Phi}$  can be also obtained by using  Audenaet's
continuity bound (\ref{Aud-CB}). But if the quantity $H_{\max}(\Phi)-H_{\min}(\Phi)$ is less then $\log(d_B-1)$ then
continuity bound (\ref{OE-CB-1}) is sharper than the continuity bound for $H_{\Phi}$ obtained via (\ref{Aud-CB}), since the functions $h_2(\varepsilon)$ and $g(\varepsilon)$ are equivalent for small $\varepsilon$.\smallskip

\begin{remark}\label{s-c-br-1} For many quantum channels $\Phi$ the quantity $H_{\max}(\Phi)-H_{\min}(\Phi)$ coincides with the
Holevo capacity of $\Phi$ that gives the ultimate rate of transmission of classical information through this channel when non-entangled
input encoding is used (in many cases it coincides with the classical capacity of $\Phi$) \cite{H-SCI,Wilde}. This holds, in particular,
for quantum channels covariant w.r.t. irreducible representation of a unitary group \cite{A,H-cov}.
\end{remark}\smallskip

\subsection{PFE-maps on the set of states with bounded energy}

By Theorem \ref{main} the output entropy $H_{\Phi}$ of any positive linear map $\Phi:\T(\H_A)\rightarrow\T(\H_B)$ preserving finiteness of the entropy
is continuous on any subset of $\S(\H_A)$, where the von Neumann entropy is continuous.

Assume that  $H_A$ is the Hamiltonian of a system $A$ with the minimal energy $E_0$ satisfying condition (\ref{H-cond}).
Then the von Neumann entropy is continuous on the set $\mathfrak{C}_{H_A,E}$ of states $\rho$ with mean energy $\Tr H_A\rho$ non exceeding $E$ (see Section 2.2). Hence  the output entropy $H_{\Phi}$ of any positive PFE-map is continuous on $\mathfrak{C}_{H_A,E}$. Moreover, it is uniformly continuous
on $\mathfrak{C}_{H_A,E}$, since this set is compact \cite{H-c-w-c}.

We will obtain uniform continuity bound for the output entropy of any PFE-map $\Phi:\T(\H_A)\rightarrow\T(\H_B)$ on the set $\mathfrak{C}_{H_A,E}$ assuming that the Hamiltonian $H_A$ satisfies the condition (\ref{H-cond+}), which is slightly stronger than condition (\ref{H-cond}), but holds for many real quantum systems. \smallskip

In the following theorem we use the notations introduced in Section 2.2. We assume that $\hat{F}_{H_A}$ is any continuous function on $\mathbb{R}_+$ satisfying conditions (\ref{F-cond-1}) and  (\ref{F-cond-2}),  $d_0$ is the minimal natural number such that  $\,\ln d_0>\hat{F}_{H_A}(0)\,$ and $\,\gamma(d)=\hat{F}^{-1}_{H_A}(\ln d)\,$ for any $d\geq d_0$.\smallskip

\begin{theorem}\label{OE-CB} \emph{Let $\,\Phi:\T(\H_A)\rightarrow \T(\H_B)$ be a  positive trace non-increasing linear map satisfying condition (\ref{PFE}) and $\,\Delta=H^\mathrm{p}_{\max}(\Phi)+1/d_0+\ln2$, where $H^\mathrm{p}_{\max}(\Phi)$ is the parameter defined in (\ref{PFE+}). Let $\bar{E}=E-E_0>0$ and $\,\varepsilon>0$. Then
\begin{equation}\label{OE-CB-3}
|H_{\Phi}(\rho)-
H_{\Phi}(\sigma)|\leq \varepsilon(1+4t)\!\left(\widehat{F}_{H_{\!A}}\!\!\left[\!\frac{\bar{E}}{(\varepsilon t)^2}\!\right]+\Delta\right)+2g(\varepsilon t)+g(\varepsilon(1+2t))
\end{equation}
for any states $\rho$ and $\sigma$ in $\,\S(\H_A)$ such that $\,\Tr H_A\rho,\Tr H_A\sigma\leq E$ and $\;\frac{1}{2}\|\shs\rho-\sigma\|_1\leq\varepsilon$ and any $\,t\in(0,T]$, where $\,T=(1/\varepsilon)\min\{1, \sqrt{\bar{E}/\gamma(d_0)}\}$ and $\,g(x)=(1+x)h_2\left(\frac{x}{1+x}\right)$.}\smallskip

\emph{If conditions (\ref{F-cond-3}) and (\ref{BD-cond}) hold~\footnote{By Proposition \ref{add-l} in Section 2.2 this holds, in particular, if  $\hat{F}_{H_A}=\hat{F}^*_{H_A}$.} then the r.h.s. of  (\ref{OE-CB-3}) can be written as
\begin{equation}\label{OE-CB-3-a}
\varepsilon(1+4t)\!\left(\ln\!\left[\!\frac{\bar{E}}{(\varepsilon t)^2}\!\right]\frac{1+o(1)}{a-1}+\mathrm{\Delta} \right)+2g(\varepsilon t)+g(\varepsilon(1+2t)),\quad\varepsilon\rightarrow0^+.
\end{equation}
and continuity bound (\ref{OE-CB-3}) with optimal $\,t$ is  asymptotically tight for large $E$.\footnote{A continuity bound $\;\displaystyle\sup_{x,y\in S_a}|f(x)-f(y)|\leq B_a(x,y)\;$ depending on a parameter $\,a\,$ is called \emph{asymptotically tight} for large $\,a\,$ if $\;\displaystyle\limsup_{a\rightarrow+\infty}\sup_{x,y\in S_a}\frac{|f(x)-f(y)|}{B_a(x,y)}=1$.}}
\end{theorem}\medskip

\begin{remark}\label{inc-p}
Since the function $\hat{F}_{H_A}$ satisfies condition (\ref{F-cond-1}) and (\ref{F-cond-2}), the r.h.s. of (\ref{OE-CB-3}) is a nondecreasing function of $\varepsilon$ and $\bar{E}$ tending to zero as $\,\varepsilon\rightarrow0^+$ for any given $\bar{E}$ and $\,t\in(0,T]$.
\end{remark}

\begin{remark}\label{t-r}
 The "free" parameter $\,t\,$ can be used to optimize continuity bound (\ref{OE-CB-3}) for given values of $E$ and $\varepsilon$.
\end{remark}
\medskip

\emph{Proof.} Note first that inequality (\ref{H-ub}) implies that
\begin{equation}\label{H-ub+}
H_{\Phi}(\rho)\leq \hat{F}_{H_A}(E(\rho)-E_0)+H^\mathrm{p}_{\max}(\Phi)
\end{equation}
for any state $\rho$ in $\S(\H_A)$ with finite $E(\rho)\doteq\Tr H_A\rho$.\smallskip

Let $\rho$ and $\sigma$ be states in $\,\S(\H_A)$ such that $\,\Tr H_A\rho,\Tr H_A\sigma\leq E$ and $\;\frac{1}{2}\|\shs\rho-\sigma\|_1\leq\varepsilon$.
By Lemma 1 in \cite{MCB} (with trivial system $B$) for any $d>d_0$ such that $\bar{E}\leq\gamma(d)$ there exist states $\varrho$, $\varsigma$, $\alpha_k$,  $\beta_k$,  $k=1,2$, in $\,\mathfrak{S}(\mathcal{H}_{A})$
and numbers $p,q\leq \displaystyle\sqrt{\bar{E}/\gamma(d)}$ such that
$\,\rank\shs \varrho,\,\rank\shs \varsigma\leq d$, $\,\mathrm{Tr}H_A \varrho, \mathrm{Tr}H_A \varsigma\leq E$, $\,\textstyle\frac{1}{2}\|\rho-\varrho\|_1\leq p$,
$\,\textstyle\frac{1}{2}\|\sigma-\varsigma\|_1\leq q$,
$\,\mathrm{Tr} \bar{H}_A\alpha_k\leq \bar{E}/p^2$, $\,\mathrm{Tr} \bar{H}_A\beta_k\leq \bar{E}/q^2$, $k=1,2$, and
\begin{equation}\label{2-r}
(1-p')\rho+p'\alpha_1=(1-p')\varrho+p'\alpha_2,\quad(1-q')\sigma+q'\beta_1
=(1-q')\varsigma+q'\beta_2,
\end{equation}
where $\bar{H}_A=H_A-E_0I_A$, $\,p'=\frac{p}{1+p}\,$ and $\,q'=\frac{q}{1+q}$. If $\,\rank\shs \rho\leq d$ we assume that $\varrho=\rho$ and do not introduce the states $\alpha_k$. Similar assumption holds if $\,\rank\shs \sigma\leq d$.

All the states $\varrho$, $\varsigma$, $\alpha_1$, $\alpha_2$, $\beta_1$ and $\beta_2$ have
finite entropy. Hence the function $H_{\Phi}$ is finite at these  states.

By using the first relation in (\ref{2-r}) and inequality (\ref{LAA}) it is easy to show that
$$
(1-p')(H_{\Phi}(\rho)-H_{\Phi}(\varrho))\leq p' (H_{\Phi}(\alpha_2)-H_{\Phi}(\alpha_1))+ h_2(p')
$$
and
$$
(1-p')(H_{\Phi}(\varrho)-H_{\Phi}(\rho))\leq p' (H_{\Phi}(\alpha_1)-H_{\Phi}(\alpha_2))+ h_2(p').
$$
These inequalities imply that
\begin{equation}\label{one}
|H_{\Phi}(\varrho)-H_{\Phi}(\rho)|\leq p|H_{\Phi}(\alpha_2)-H_{\Phi}(\alpha_1)|+ g(p).
\end{equation}
Similarly, by using the second relation in (\ref{2-r}) and inequality (\ref{LAA}) we obtain
\begin{equation}\label{two}
|H_{\Phi}(\varsigma)-H_{\Phi}(\sigma)|\leq q|H_{\Phi}(\beta_2)-H_{\Phi}(\beta_1)|+ g(q).
\end{equation}

Since $\,\mathrm{Tr} \bar{H}_A\alpha_k\leq \bar{E}/p^2$ and  $\,\mathrm{Tr} \bar{H}_A\beta_k\leq \bar{E}/q^2$ , $k=1,2$, it follows from
(\ref{H-ub+}) that
\begin{equation}\label{one+}
|H_{\Phi}(\alpha_2)-H_{\Phi}(\alpha_1)|\leq\widehat{F}_{H_A}\!\left(\bar{E}/p^2\right)+H^\mathrm{p}_{\max}(\Phi)
\end{equation}
and
\begin{equation}\label{two+}
|H_{\Phi}(\beta_2)-H_{\Phi}(\beta_1)|\leq\widehat{F}_{H_A}\!\left(\bar{E}/q^2\right)+H^\mathrm{p}_{\max}(\Phi).
\end{equation}
Since $p,q\leq y\doteq\displaystyle\sqrt{\bar{E}/\gamma(d)}\,$ and the function $\,E\mapsto\widehat{F}_{H_A}(E)/\sqrt{E}\,$ is non-increasing, we have
$$
x\widehat{F}_{H_A}\!\left(\bar{E}/x^2\right)\leq y\widehat{F}_{H_A}\!\left(\bar{E}/y^2\right)=\sqrt{\bar{E}/\gamma(d)}\,\widehat{F}_{H_A}\!\left(\gamma(d)\right)=
\sqrt{\bar{E}/\gamma(d)}\ln d,
$$
$x=p,q$, where the last equality follows from the definition of $\gamma(d)$.

Thus, it follows from (\ref{one})-(\ref{two+}) and the monotonicity of the function $g(x)$ that
\begin{equation}\label{three+}
\!|H_{\Phi}(\varrho)-H_{\Phi}(\rho)|,|H_{\Phi}(\varsigma)-H_{\Phi}(\sigma)|\leq \shs\sqrt{\bar{E}/\gamma(d)}\left(\ln d+H^\mathrm{p}_{\max}(\Phi)\right)+g\!\left(\sqrt{\bar{E}/\gamma(d)}\right).
\end{equation}

Since $\,\rank\shs \varrho\leq d$ and $\,\rank\shs \varsigma\leq d$, the supports of both states
$\varrho$ and $\varsigma$ are contained in some $2d$-dimensional subspace of $\H_A$. By the triangle inequality
we have
$$
\|\varrho-\varsigma\|_1\leq \|\varrho-\rho\|_1+\|\varsigma-\sigma\|_1+\|\rho-\sigma\|_1\leq 2\varepsilon+4\sqrt{\bar{E}/\gamma(d)}.
$$
So, by using the Alicki-Fannes-Winter method (mentioned in Section 5.1) and inequality (\ref{H-ub}) one can show that
\begin{equation}\label{fcb-c}
 |H_{\Phi}(\varrho)-H_{\Phi}(\varsigma)|\leq \left(2\sqrt{\bar{E}/\gamma(d)}+\varepsilon\right) (\ln (2d)+H^\mathrm{p}_{\max}(\Phi))+g\!\left(2\sqrt{\bar{E}/\gamma(d)}+\varepsilon\right).
\end{equation}

It follows from (\ref{three+}) and (\ref{fcb-c}) that
\begin{equation}\label{m-cb}
\begin{array}{c}
|H_{\Phi}(\rho)-H_{\Phi}(\sigma)|\,\leq   \displaystyle \left(4\sqrt{\bar{E}/\gamma(d)}+\varepsilon\right) (\ln d+H^\mathrm{p}_{\max}(\Phi))\\\\ +  \displaystyle\left(2\sqrt{\bar{E}/\gamma(d)}+\varepsilon\right)\ln 2+g\!\left(2\sqrt{\bar{E}/\gamma(d)}+\varepsilon\right)+2 g\!\left(\sqrt{\bar{E}/\gamma(d)}\right).
\end{array}
\end{equation}

If $t\in(0,T]$ then, since the sequence $\gamma(d)$ is increasing, there is a natural number $d_*>d_0$  such that $\gamma(d_*)>\bar{E}/(\varepsilon t)^2\geq \bar{E}$ but $\gamma(d_*-1)\leq \bar{E}/(\varepsilon t)^2$. It follows that
$$
\sqrt{\bar{E}/\gamma(d_*)}\leq \varepsilon t\leq 1\quad \textrm{and} \quad
\ln (d_*-1) = \widehat{F}_{H_A}(\gamma(d_*-1))\leq \widehat{F}_{H_A}(\bar{E}/(\varepsilon t)^2),
$$
where the first condition in (\ref{F-cond-2}) was used.
Since $\ln d_*\leq\ln (d_*-1)+1/(d_*-1)\leq\ln (d_*-1)+1/d_0$, inequality (\ref{m-cb}) with $d=d_*$ implies continuity bound (\ref{OE-CB-3}).

If conditions (\ref{F-cond-3}) and (\ref{BD-cond}) hold then part B of Proposition \ref{add-l}
shows that $\hat{F}_{H_A}(E)=(a-1)^{-1}\ln(E)(1+o(1))$ as $E\to+\infty$.
This implies the asymptotic representation (\ref{OE-CB-3-a}) of the r.h.s. of (\ref{OE-CB-3}). The asymptotic tightness of
continuity bound (\ref{OE-CB-3}) in this case follows from the asymptotic tightness of the
continuity bound for the von Neumann entropy presented in \cite[Example 2]{MCB}, since the right hand sides of these continuity bounds
coincide provided that $H^\mathrm{p}_{\max}(\Phi)=0$. $\square$ \medskip

Assume now that the input system $A$ is the $\,\ell$-mode quantum oscillator with the frequencies $\,\omega_1,...,\omega_{\ell}\,$. The Hamiltonian of this system has the form
\begin{equation*}%\label{qos-H}
H_A=\sum_{i=1}^{\ell}\hbar \omega_i a_i^*a_i+E_0 I_A,\quad E_0=\frac{1}{2}\sum_{i=1}^{\ell}\hbar \omega_i,
\end{equation*}
where $a_i$ and $a^*_i$ are the annihilation and creation operators of the $i$-th mode \cite{H-SCI}. Note that this
Hamiltonian satisfies condition (\ref{BD-cond}) with $a=1+1/\ell$ \cite{Datta}.

In this case the function
\begin{equation*}%\label{F-ub+}
\bar{F}_{\ell,\omega}(E)\doteq \ell\ln \frac{E+2E_0}{\ell E_*}+\ell,\quad E_*=\left[\prod_{i=1}^{\ell}\hbar\omega_i\right]^{1/\ell},%\vspace{-5pt}
\end{equation*}
is an  upper bound on the function $\bar{F}_{H_A}(E)\doteq F_{H_A}(E+E_0)$  satisfying all the conditions in (\ref{F-cond-1}),(\ref{F-cond-2}) and (\ref{F-cond-3}) \cite{MCB}. By using the function $\bar{F}_{\ell,\omega}$ in the role of function $\hat{F}_{H_A}$
in Theorem \ref{OE-CB} we obtain the following\smallskip\pagebreak

\begin{corollary}\label{OE-CB-G-1}
\emph{Let $\,\Phi:\T(\H_A)\rightarrow \T(\H_B)$ be a  positive trace non-increasing linear map satisfying condition (\ref{PFE}), where $A$ is the $\ell$-mode quantum oscillator with the frequencies $\omega_1,...,\omega_{\ell}$. Let
$\bar{E}=E-E_0>0$, $\varepsilon>0$ and $\,T_*=(1/\varepsilon)\min\{1, \sqrt{\bar{E}/E_0}\}$. Then
\begin{equation}\label{OE-ineq+}
\begin{array}{c}
\displaystyle |H_{\Phi}(\rho)-
H_{\Phi}(\sigma)|\leq \varepsilon(1+4t)\left(\ell\ln \frac{\bar{E}/(\varepsilon t)^2+2E_0}{\ell E_*}+\ell+\mathrm{\Delta}^*\!\right)+2g(\varepsilon t)+g(\varepsilon(1+2t))
\end{array}\!\!
\end{equation}
for any states $\rho$ and $\sigma$ in $\,\S(\H_A)$ such that $\,\Tr H_A\rho\leq E,\,\Tr H_A\sigma\leq E$ and $\;\frac{1}{2}\|\shs\rho-\sigma\|_1\leq\varepsilon$ and any $\,t\in(0,T_*]$, where $\,\mathrm{\Delta}^*=H^\mathrm{p}_{\max}(\Phi)+e^{-\ell}+\ln2$  (the parameter $H^\mathrm{p}_{\max}(\Phi)$ is defined in (\ref{PFE+})).}\smallskip

\emph{Continuity bound (\ref{OE-ineq+}) with optimal $\,t$ is  asymptotically tight for large $E$.}
\end{corollary}\smallskip

\emph{Proof.} All the assertions of the corollary directly follow from Theorem \ref{OE-CB}. It suffices to note that in this case
$d_0$ is the minimal natural number not less than $x^{\ell}$, where $x=2E_0e/(\ell E_*)\geq e$, and hence
$$
\gamma(d_0)\doteq\bar{F}^{-1}_{\ell,\omega}(\ln d_0)=(\ell/e)E_*\sqrt[\ell]{d_0}-2E_0\leq(\ell/e)E_*x\sqrt[\ell]{1+e^{-\ell}}-2E_0\leq E_0.\;\;\square
$$

In the case $\Phi=\id_A$ continuity bounds (\ref{OE-CB-3}) and (\ref{OE-ineq+}) coincide with the
continuity bounds for the von Neumann entropy under the energy constraint obtained in \cite{MCB}.
%\bigskip

%{\bf Acknowledgments.}  I am grateful to A.S.Holevo for useful remarks.
%\medskip

\end{document}